\documentclass[twocolumn]{aastex62}

\usepackage{enumitem}

\newcommand{\TESS}{\emph{TESS}}
\newcommand{\TEVSS}{\emph{TE$_\mathrm{v}$SS}}

\newcommand{\Rsun}{\mbox{$R_{\odot}$}}
\newcommand{\rearth}{R$_{\oplus}$}

\newcommand{\allesfitter}{\texttt{allesfitter}}

\begin{document}

\title[Searching for Space Vampires with {\TEVSS}]{Searching for Space Vampires with {\TEVSS}}
\shorttitle{Searching for Space Vampires with {\TEVSS}}

\correspondingauthor{Maximilian N.\ G{\"u}nther}
\email{maxgue@mit.edu}

\author[0000-0002-3164-9086]{Maximilian N.\ G{\"u}nther}
\affil{Department of Physics, and Kavli Institute for Astrophysics and Space Research, Massachusetts Institute of Technology, Cambridge, MA 02139, USA}
\affil{Juan Carlos Torres Fellow}

\author[0000-0001-6298-412X]{David A.\ Berardo}
\affil{Department of Physics, and Kavli Institute for Astrophysics and Space Research, Massachusetts Institute of Technology, Cambridge, MA 02139, USA}

\begin{abstract}
%Context/punchline
It is a truth universally acknowledged, that a single human in possession of a good space telescope, must be in search of a space vampire.
%Aim
Here, we showcase our search for transit signatures of tidally locked space vampires, trapped in the gravitational pull of late M-dwarfs.
%Methods
We generate forward models representing two potential space vampire populations -- those in bat shape and those in humanoid shape. We search lightcurves from the Transiting Exo-Vampire Survey Satellite {\TEVSS} using a template matching algorithm and fit them using our \texttt{allesfitter} software.
%Results
Adding the information gained from {\TEVSS} data, we greatly decrease uncertainty for the existence and occurrence rates of space vampires, and constrain $\eta_\mathrm{Space Vampire}$ to a range of 0\% to 100\% (or more).
%Impact
These precise analyses will be crucial for optimizing future observing schedules for space-vampire characterization with the James Webb Space-Vampire Telescope (\textit{JWS$_\mathrm{V}$T}) and the Extremely-Large-Vampire Telescopes (\textit{EL$_\mathrm{V}$Ts}).
\end{abstract}


%%%%%%%%%%%%%%%%%%%%%%%%%%%%%%%%%%%%%%%%%%%%%%%%%%%%%%%%%%%%%%%%%%%%%%%%%%
% INTRODUCTION
%%%%%%%%%%%%%%%%%%%%%%%%%%%%%%%%%%%%%%%%%%%%%%%%%%%%%%%%%%%%%%%%%%%%%%%%%%

\section{Introduction}
\label{s:Introduction}
Space vampire detection and characterization is a major goal of research, helping us to understand space vampires' sizes, masses, composition, birth and migration histories.
While \cite{Stoker1897} hypothesized that vampires may only be Earth-bound and predominantly originate in Transylvania, more recent theories predicted their origin to be connected to outer space \citep[e.g.][]{OBrien1975}. %Rocky Horror Picture Show
This led to the \textit{vampanspermia} theory \citep{Liter-Allyno2004},
%2014: What we do in the shadows
postulating that vampires might have originated in outer space, settled down and `domesticated' asteroids, and then fell onto Earth through meteor crashes.
The outer space origin, so these authors claim, would also explain their fear of solar light.
Said theory led to plenty of controversy in the community, splitting scientists into opposing groups: those looking for peaceful communication, and those demanding to increase the garlic concentration in Earth's atmosphere.

\begin{figure}
    \centering
    \includegraphics[width=0.7\columnwidth]{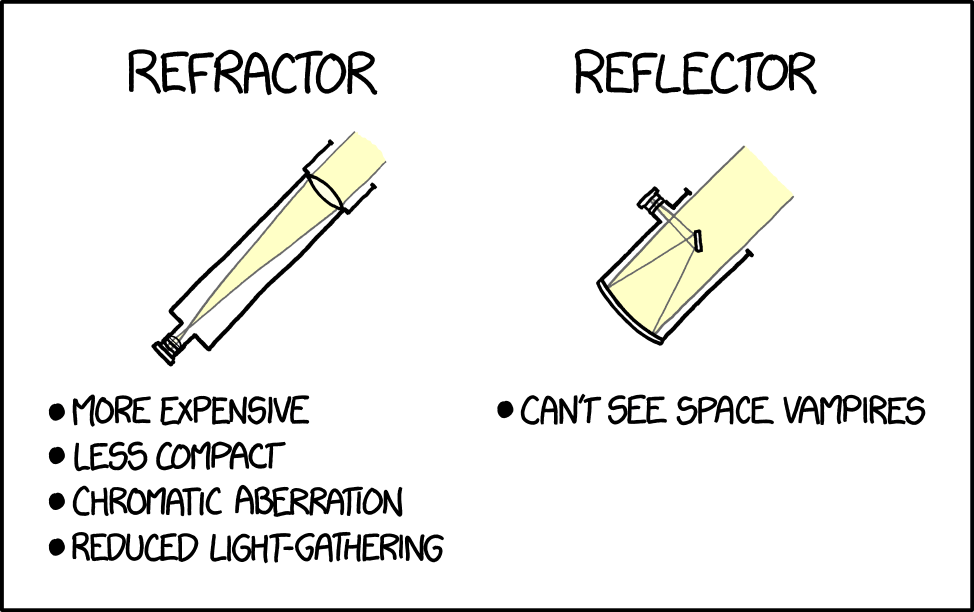}
    \caption{{\TEVSS} is equipped with refractor telescopes rather than reflector telescopes. This serves only one purpose: to allow the efficient detection of space vampires. Illustration: Randall Munroe. Source: \url{https://xkcd.com/1791/}}
    \label{fig:xkcd}
\end{figure}

The Transiting Exo-Vampire Survey Satellite ({\TEVSS})\footnote{often mistakenly referred to as the Transiting Exoplanet Survey Satellite ({\TESS}) -- but we all know its true purpose...} was launched in 2018 with the goal of detecting space-vampires transiting in front of bright stars in our galactic neighborhood.
{\TEVSS} is equipped with four red-sensitive CCD cameras that span a total field of view of 96 x 21 square degrees.
Every 28 days, {\TEVSS} shifts its field of view, progressively scanning through the entire sky in its two year primary mission.
Most importantly, {\TEVSS} is equipped with refractor telescopes rather than reflector telescopes. This is to avoid the usage of mirrors, which would render the detection of space vampires impossible (Fig.~\ref{fig:xkcd}).

We identify three possible scenarios that could enable space vampire identification:
\begin{enumerate}
    \item Tidally locked space vampires around late M-dwarfs. These creatures might have originated either free floating or on interstellar asteroids, and then got trapped in the gravitational pull of a late M-dwarf. This is not paradoxical to their fear of solar light. It is known that Earth-bound vampires cannot withstand solar light (5700\,K), but are unaffected by candle light and chimney fires (1900\,K). In a series of laboratory biology experiments, Earth-bound vampire cells were found resistant to blackbody spectra up to 2800\,K, with a sudden drop off for higher temperatures \citep{Schreck1922}.
    In terms of stellar types, and assuming these laboratory results translate into the space vampire regime, this means space vampires should be able to withstand the radiation from M6V dwarfs and later stellar types.
    \item Free-floating space vampires which either inhabit or cross our solar system. In this case the danger for Earth is imminent, and we do not wish to be responsible for any right or wrong predictions. We thus say ``this is beyond the scope of this paper'' and leave it to solar system scientists to figure out. In fact, a second paper (Van Helsing et al., in prep.) will dwell into free-floating space vampires, a search for their transients in our Earth neighborhood, and necessary countermeasures. Be prepared.
    \item Space vampires are already among us. They can be easily detected by a sudden feeling of being bitten in the neck, followed by a blood rush to the head and an immediate feeling of drowsiness; in which case it is obviously too late.
\end{enumerate}

We dedicate this paper to case 1, the study of tidally locked space vampires around late M-dwarfs. In section~\ref{s:Methods} we explain the forward model we developed to create transit shapes from both bat- and humanoid-shaped vampires. We then avoid going into details on our cross-matching search for these, and quickly present our results in section~\ref{s:Results}. Finally, we discuss and conclude our study in sections \ref{s:Discussion} and \ref{s:Conclusion}.

% Vampires aren't real

%%%%%%%%%%%%%%%%%%%%%%%%%%%%%%%%%%%%%%%%%%%%%%%%%%%%%%%%%%%%%%%%%%%%%%%%%%
% METHODS
%%%%%%%%%%%%%%%%%%%%%%%%%%%%%%%%%%%%%%%%%%%%%%%%%%%%%%%%%%%%%%%%%%%%%%%%%%
\section{Methods}
\label{s:Methods}

\begin{figure}
    \centering
    \includegraphics[width=\columnwidth]{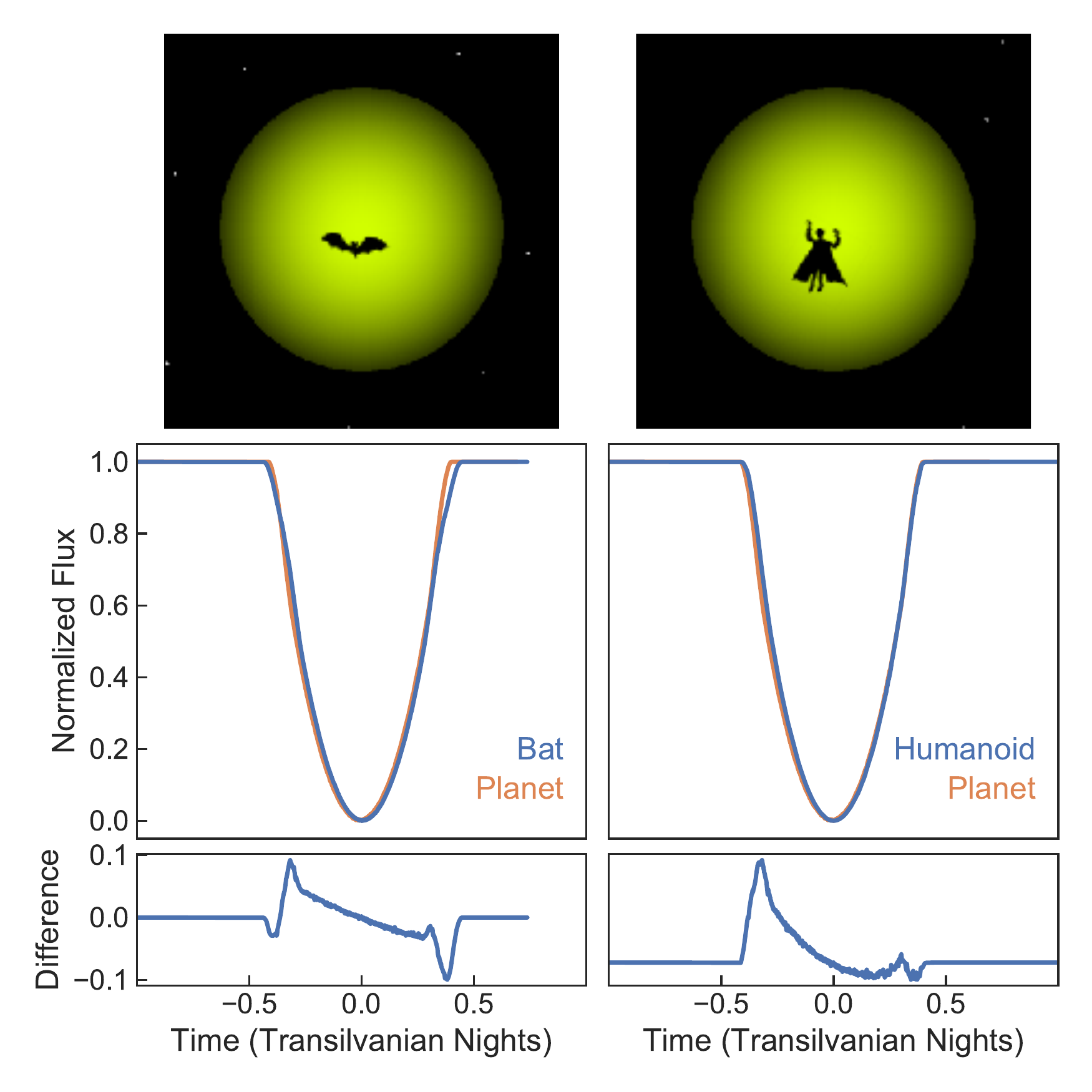}
    \caption{Transit shapes for a bat (left column), humanoid vampire (right column) and planet in comparison. 
    The top panels shows our forward modeling, which generates lightcurves by integrating over the visible stellar surface.
    The middle panels shows the resulting, normalized transit shapes, with depths normalized to a range of 0 to 1, and durations normalized to 1 Transylvanian Night. 
    The bottom panels shows the difference lightcurve between bat and planet (left), and humanoid vampire and planet (right).
    This comparison highlights the clear distinction between bats, humanoid vampires and planets according to their transit shape.
    Note that, theoretically, their different sizes could be used as an alternative distinction method.}
    \label{fig:transit_model}
\end{figure}

First, we created a foreward model of space vampire transits. We manually drew the shape of a bat and a humanoid vampire in the software package \texttt{paint}, and then used \texttt{processing}\footnote{https://processing.org/} (a Java-based IDE)  to pass this shape in front of a star with a quadratic limb-darkening model. At each time step, we integrated the brightness of the visible stellar surface over an 800x400 grid of points, creating a lightcurve with the distinct transit shape of a bat or a humanoid vampire (Fig.~\ref{fig:transit_model}). 
Additionally, we repeated this process to generate the model of a transiting planet.
We then compared the normalized transit shapes, with depths normalized to a range of 0 to 1, and durations normalized to 1 Transylvanian Night. 
We find that analyzing the normalized transit shapes enables a clear and unique distinction between bats, humanoid vampires and planets.
We note that, in theory, the difference in size (planets usually range from 0.8--22\,\rearth, while bats and humanoid vampires are typically around 0.1--2\,m) could also be used to differentiate the signals (beyond the scope of this work).

With this template, we then iterated over all dwarf stars later than M6-type observed in {\TEVSS} 2 minute cadence, whose lightcurves were processed with the \texttt{bite} pipeline \citep{TotaLlymadeup1992}.
For each target star, we scaled the templates' signatures down to account for the appropriate size ratio between the bat ($\sim$0.1\,m), the humanoid vampire ($\sim$1.8\,m), and the target star ($<$0.127\,\Rsun{}).
This led to transit depths of $\sim1.3$\,ppq for the bats and $\sim415$\,ppq for the humanoids\footnote{ppq: parts-per-quintrillion, $10^{-18}$}.
However, detecting a transit signal with such a depth is not enough; it could also stem from a really, really, really small planet, potentially one with an Earth-like atmosphere, surface water and extraterrestrial life. 
To avoid confusing space vampires with habitable micro-planets, we thus emphasize that in order to identify a space vampire, one must pay attention to their transit shape and deviations from the planetary transit shape.
As discussed above, this leads to a requirement of measuring the transit shapes to a photometric precision of at least 10\% of the depth (see Fig.~\ref{fig:transit_model}), leading to $\sim0.1$\,ppq for the bats and $\sim42$\,ppq for the humanoids.

As a self-less service to the community, we implemented the new bat and humanoid vampire transit shapes as modules in our \allesfitter{} software \citep[][and in prep.]{allesfitter}\footnote{\url{https://github.com/MNGuenther/allesfitter}} and performed a homogeneous analysis of all photometric \TEVSS{} data. 
\allesfitter{} is a convenient wrapper around any packages needed for space vampire research and garlic-heavy cooking recipes.
All input parameters and settings can be defined in a graphical or programmatic user interface, and then the code automatically runs a nested sampling or MCMC fit. It not only produces all output such as tables, latex tables, and plots, but also shares a selfie of you on \textit{vampagram} and \textit{biter}.

% Ouch, I woke up, my neck was swollen and I had two stings that looked like bite marks...

%%%%%%%%%%%%%%%%%%%%%%%%%%%%%%%%%%%%%%%%%%%%%%%%%%%%%%%%%%%%%%%%%%%%%%%%%%
% RESULTS
%%%%%%%%%%%%%%%%%%%%%%%%%%%%%%%%%%%%%%%%%%%%%%%%%%%%%%%%%%%%%%%%%%%%%%%%%%
\section{Results}
\label{s:Results}

Upon inspecting all candidate threshold events by eye, we identified a short list of between 0 and 394400933 potential space vampire transits. 
%0-394-40093-3 ISBN of "Space Vampires"
Using Bayesian evidence, we determined that two of these most likely originated from bats (or noise), while one was more likely due to humanoid shapes (or noise). The remaining 0 to 394400930 could be due to either shape (or noise).
There is a small possibility, which we cannot yet confidently rule out, that some of these signals might be noise features rather than space vampire transits.
After all, the noise floor of {\TEVSS} per 2 minute exposure is a factor of circa $10^{15}$ higher than the expected signals.

% I can't stand the light anymore... why do my teeth feel so sharp...

%%%%%%%%%%%%%%%%%%%%%%%%%%%%%%%%%%%%%%%%%%%%%%%%%%%%%%%%%%%%%%%%%%%%%%%%%%
% DISCUSSION
%%%%%%%%%%%%%%%%%%%%%%%%%%%%%%%%%%%%%%%%%%%%%%%%%%%%%%%%%%%%%%%%%%%%%%%%%%
\section{Discussion}
\label{s:Discussion}

Taking into account the biases in target selection and observing patterns, as well as {\TEVSS} detection efficiency in the parts-per-quintrillion regime, we estimate that the occurence rate of space vampires around late M dwarfs lies between 0\% and 100\% (or more).
After centuries of wondering and not knowing things \citep{Plato400, Snow299}, we consider this result a major breakthrough, while also pointing out that much more work has to be done.
Future dedicated missions such as the James Webb Space-Vampire Telescope (\textit{JWS$_\mathrm{V}$T}) and the Extremely-Large-Vampire Telescopes (\textit{EL$_\mathrm{V}$Ts}) will play a pivotal role in narrowing the error bars on our measurements, and studying the amount of $C_2 H_5 O H$ in vampires' breath.

% Why is my mirror image washing out...

%%%%%%%%%%%%%%%%%%%%%%%%%%%%%%%%%%%%%%%%%%%%%%%%%%%%%%%%%%%%%%%%%%%%%%%%%%
% CONCLUSION
%%%%%%%%%%%%%%%%%%%%%%%%%%%%%%%%%%%%%%%%%%%%%%%%%%%%%%%%%%%%%%%%%%%%%%%%%%
\section{Conclusion}
\label{s:Conclusion}

We developed models for space vampire transits on cool dwarf stars, and pretended to have conducted an extensive search for their signals. We showed that the space vampire occurence rates lie between 0\% and 100\% (or more). It is comforting to know that it cannot be less than that (albeit possibly more than that), and we consider this a major break through on our side. More funding is welcome.

% I feel the need to bite...

%%%%%%%%%%%%%%%%%%%%%%%%%%%%%%%%%%%%%%%%%%%%%%%%%%%%%%%%%%%%%%%%%%%%%%%%%%
% Acknowledgments
%%%%%%%%%%%%%%%%%%%%%%%%%%%%%%%%%%%%%%%%%%%%%%%%%%%%%%%%%%%%%%%%%%%%%%%%%%
\section*{Acknowledgments}
\label{s:Acknowledgments}
MNG and DAB are in no way supported by the Space Vampire Research Council (SVRC) grant number 1897-05-26.
%1897 May 26: date of Stoker's book release

%"What are we gonna do today, Brain?" "Same thing we do everyday, Pinky, stay home and flatten curve!"

{\scriptsize
\bibliographystyle{apalike}
\bibliography{fake_references.bib}
}

\end{document}